\documentstyle[12pt]{article}
\newcommand{\agt}{\mathrel{\raisebox{-.6ex}{$\stackrel{\textstyle>}{\sim}$}}}
\def\overlay#1#2{\ifmmode \setbox 0=\hbox {$#1$}\setbox 1=\hbox to\wd 0{\hss
$#2$\hss }\else \setbox 0=\hbox {#1}\setbox 1=\hbox to\wd 0{\hss #2\hss }\fi
#1\hskip -\wd 0\box 1}
\def\case#1/#2{{\textstyle{#1\over#2}}}

\def\etal{{\it et al.}}

\catcode`@=11
\newcount\@tempcntc
\def\@citex[#1]#2{\if@filesw\immediate\write\@auxout{\string\citation{#2}}\fi
  \@tempcnta\z@\@tempcntb\m@ne\def\@citea{}\@cite{\@for\@citeb:=#2\do
    {\@ifundefined
      {b@\@citeb}{\@citeo\@tempcntb\m@ne\@citea\def\@citea{,}{\bf ?}\@warning
      {Citation `\@citeb' on page \thepage \space undefined}}%
    {\setbox\z@\hbox{\global\@tempcntc0\csname b@\@citeb\endcsname\relax}%
     \ifnum\@tempcntc=\z@ \@citeo\@tempcntb\m@ne
       \@citea\def\@citea{,}\hbox{\csname b@\@citeb\endcsname}%
     \else
      \advance\@tempcntb\@ne
      \ifnum\@tempcntb=\@tempcntc
      \else\advance\@tempcntb\m@ne\@citeo
      \@tempcnta\@tempcntc\@tempcntb\@tempcntc\fi\fi}}\@citeo}{#1}}
\def\@citeo{\ifnum\@tempcnta>\@tempcntb\else\@citea\def\@citea{,}%
 \ifnum\@tempcnta=\@tempcntb\the\@tempcnta\else
  {\advance\@tempcnta\@ne\ifnum\@tempcnta=\@tempcntb \else \def\@citea{--}\fi
   \advance\@tempcnta\m@ne\the\@tempcnta\@citea\the\@tempcntb}\fi\fi}
\catcode`@=12

\font\fortssbx=cmssbx10 scaled \magstep2

\begin{document}

\thispagestyle{empty}

\hbox to \hsize{
\hskip.5in \raise.1in\hbox{\fortssbx University of Wisconsin - Madison}
\hfill  $\vcenter{\hbox{\bf MADPH-95-890}\vskip-.3cm
                   \hbox{\bf UICHEP-TH/95-4}\vskip-.3cm
                   \hbox{\bf RAL-TR-95-015}\vskip-.3cm
                   \hbox{\bf hep-ph/9506321}\vskip-.3cm
                   \hbox{June 1995}}$ }

\vspace{.75in}

\begin{center}
{\large\bf Possible pre-LEP200 SUSY threshold signals}\\[.4in]
V.~Barger$^a$, W.-Y.~Keung$^b$ and R.J.N.~Phillips$^c$\\[.2in]
\it
$^a$Physics Department, University of Wisconsin, Madison, WI 53706, USA\\
$^b$Physics Department, University of Illinois at Chicago, IL 60607-7059, USA\\
$^c$Rutherford Appleton Laboratory, Chilton, Didcot, Oxon OX11 0QX, UK
\end{center}

\vspace{.5in}

\begin{abstract}
If $R$-parity is not conserved, the SUSY-threshold production process
$e^+e^- \to \chi_1^0\chi_1^0$ could be detectable with relatively low
luminosity.
Hence an interesting mass range for the lightest SUSY particle $\chi_1^0$
could be explored at the CERN LEP collider during its intermediate
energy development, even before the full LEP200 upgrade is completed.
We present cross section formulas and discuss event rates and detection
for the three distinct decay options:
$\chi_1^0\to 2\, {\rm charged \; leptons}\, +\, {\rm neutrino}$,
$\chi_1^0\to {\rm lepton}\, +2\, {\rm quarks}$, and
$\chi_1^0\to 3\, {\rm quarks}$.
\end{abstract}

\newpage
The theoretical attractions of Supersymmetry (SUSY) still lack the
direct experimental support that would come from discovering some of
the predicted SUSY partners of Standard Model (SM) particles.  Their
non-appearance may be attributed to their masses being beyond the
reach of experimental searches so far. Also, in the Minimal
Supersymmetric Standard Model (MSSM)\cite{mssm}, the lightest SUSY
partner (usually expected to be the lightest neutralino $\chi_1^0$)
is stable and weakly interacting so that the lowest-threshold SUSY
process
\begin{equation}
          e^+e^-\to\chi_1^0\chi_1^0
\label{eq:prod}
\end{equation}
is completely invisible.  But if $\chi_1^0$ in fact decays visibly,
as can happen if the conventional assumption of $R$-parity
conservation is relaxed\cite{rpv,dawson,bgh,emr,dr}, this process will be
detectable and in fact may have an appreciable cross section, depending
on the mixture of electroweak gauginos and higgsinos in $\chi_1^0$
and on the mass of the lightest selectron.
(Strictly speaking,  $R$-parity violation (RPV) also allows
single-SUSY-partner production with lower thresholds than
Eq.(\ref{eq:prod}), but the corresponding production cross
sections depend on unknown RPV-couplings that are either
constrained\cite{bgh,emr,grt} or suspected to be much smaller than the
gauge and Yukawa couplings controlling Eq.(\ref{eq:prod})).
In the present Letter we provide cross section formulas
and numerical evaluations for Eq.(\ref{eq:prod}), and discuss
the types of signal that would be observable in an RPV scenario
where $\chi_1^0$ is the lightest SUSY partner (LSP).   We also
point out that these cross sections are big enough to enable
significant SUSY-RPV searches during the gradual upgrade of
the CERN $e^+e^-$ collider LEP from CM energy $\sqrt s \simeq 90$ GeV
toward $200$ GeV, even though the intermediate-energy running will
not accumulate high luminosity.   An experimental search has already
been made for Eq.(\ref{eq:prod}) at LEP\cite{opal}, assuming that
$\chi_1^0$ is a pure photino and taking a specific RPV model\cite{babu};
it excludes a range of photino masses between 5 and 42 GeV in this model,
provided the exchanged selectron is light enough.  Also possible LEP200
$\tau$-lepton signals from Eq.(1) have been considered in Ref.\cite{grt}.
But other discussions of SUSY-RPV signals at $e^+e^-$ colliders
\cite{rpv,dawson,dl} usually neglect Eq.(\ref{eq:prod}) in favor of
processes such as slepton-pair or chargino-pair production, which may
have larger cross sections eventually but also have higher thresholds.

With the MSSM particle content, the most general gauge- and
SUSY-invariant Lagrangian includes the following Yukawa coupling
terms\cite{rpv,dawson}:
\begin{equation}
{\cal L}_{RPV} = \lambda_{ijk} L_i L_j E_k^c
                +\lambda_{ijk}^\prime L_i Q_j D_k^c
                +\lambda_{ijk}^{\prime \prime} U_i^c D_j^c D_k^c,
\label{eq:rpv}
\end{equation}
where $L_i$ and $E_i^c$ are the (left-handed)  lepton doublet and
antilepton singlet chiral superfields (with generation index $i$),
while $Q_i$ and $U_i^c$, $D_i^c$ are the quark doublet and
antiquark singlet superfields.  Antisymmetry gives $\lambda_{ijk}=-
\lambda_{jik}$ and $\lambda_{ijk}^{\prime\prime}=-\lambda_{ikj}^
{\prime\prime}$.  The $LLE^c$ and $LQD^c$ terms
violate lepton number $L$ while the $U^cD^cD^c$ terms violate
baryon number $B$.  In the MSSM these terms are all conventionally
forbidden by a multiplicative symmetry called $R$-parity ($R_p$),
with $R_p=1$ for all SM particles and $R_p=-1$ for their SUSY partners,
in order to prevent rapid proton decay.  However, proton decay is
prevented if {\it either} the $L$-violating
{\it or} the $B$-violating terms are absent; this is
an adequate restriction on RPV scenarios.  Since $\chi_1^0$ can
couple to any of these superfields, each RPV term provides a
possible decay channel into SM fermions (via sfermion exchanges)
as follows:
\begin{eqnarray}
\lambda_{ijk}
&\Rightarrow&\chi_1^0\to\ell_i^-\nu_j\ell_k^+,\label{eq:dec1}\\
\lambda_{ijk}^\prime&\Rightarrow&\chi_1^0\to\ell_i^-u_j\bar d_k,\;
                                   \nu_i   d_j\bar d_k,\label{eq:dec2}\\
\lambda_{ijk}^{\prime\prime}&\Rightarrow&\chi_1^0\to\bar u_i\bar d_j \bar d_k,
\label{eq:dec3}
\end{eqnarray}
together with the charge-conjugate channels.  In practical applications,
it is customary to consider just one of the 45 independent couplings
in Eq.(\ref{eq:rpv}) at a time, since it seems likely that one will
be more important than the others. For example,  the OPAL search\cite{opal}
probed $\chi_1^0 \to e\nu_\mu\tau, \nu_e\mu\tau$ signals\cite{babu} due
to $\lambda_{123}$.  (More complicated situations, where several
$\lambda_{ijk},\; \lambda_{lmn}^{\prime}$ contributions are comparable,
are discussed however in Ref.\cite{grt}).   The requirement that $\chi_1^0$
decays within the detector (typically within 1\,m) translates
into\cite{dawson}
\begin{equation}
\lambda \agt 5\times 10^{-6}\sqrt {\beta\gamma}\; (\tilde m_f)^2 \;
                                        (m_{\chi_1^0})^{-5/2},
\label{eq:det}
\end{equation}
where $\lambda$ denotes the dominant RPV coupling, $m_{\chi_1^0}$ and
$\tilde m_f$ are the masses of $\chi_1^0$ and the dominant exchanged
sfermion (in GeV), while $\beta\gamma=\sqrt {s/(4m_{\chi_1^0}^2)-1}$
is the appropriate Lorentz factor for $\chi_1^0$.  For typical values
of present interest $m_{\chi_1^0} \sim 50$, $\tilde m_f\sim 100$,
$\beta\gamma\sim 1$, this gives a very weak lower bound
$\lambda\agt 3\times 10^{-6}$.
The alternative types of decay mode in Eqs.(\ref{eq:dec1})--(\ref{eq:dec3})
give quite different final state decay signatures for Eq.(\ref{eq:prod}),
which we now discuss.\\
(a) $LLE^c$-mediated decays: Eq.(\ref{eq:dec1}). Each $\chi_1^0$ decays to
two charged plus one neutral lepton; e.g. the $\lambda_{132}$ mode gives
\begin{equation}
\chi_1^0\to e^-\nu_{\tau}\mu^+,\; \nu_e\tau^-\mu^+,\; e^+\bar\nu_{\tau}
\mu^-,\; \bar\nu_e\tau^+\mu^- ,
\label{eq:lle}
\end{equation}
with equal probabilities, assuming the contributing sfermion masses do
not depend on the generation.  In this case
all final states contain two muons; $50\%$ contain same-sign dimuons;
$12.5\%$ contain same-sign dimuons plus same-sign dielectrons. In other
cases the lepton flavors are distributed differently, but all have missing
energy-momentum and same-sign dileptons among their signatures.
Because of the missing neutrinos here, one cannot directly reconstruct
the $\chi_1^0$ mass, but it can in principle be inferred from the
distribution of opposite-sign-dilepton invariant mass $m(\ell^+\ell^-)$
(with four entries per event, two of which have a common $\chi_1^0$
parent and hence $m < m_{\chi_1^0}$).\\
(b) $LQD^c$-mediated decays: Eq.(\ref{eq:dec2}). Each $\chi_1^0$ decays to
a charged or neutral lepton plus two quarks (i.e. potentially two jets);
e.g.\ the $\lambda_{213}^{\prime}$ mode gives
\begin{equation}
\chi_1^0\to \mu^- u \bar b,\;
\nu_\mu d \bar b,\; \mu^+\bar ub,\; \bar\nu_\mu \bar d b \ ,
\label{eq:lqd}
\end{equation}
with comparable probabilities, that become equal if $\tilde m_{\mu L}=
\tilde m_{\nu_\mu L},\;\tilde m_{uL}=\tilde m_{dL}$ and $\chi_1^0$
is a pure bino.   In the latter case, $50\%$ of final states have dimuons
(plus jets plus no missing energy), $25\%$ have same-sign dimuons.
We can attempt to reconstruct the $\chi_1^0$  mass as follows.
We partition each $\ell\ell jjjj$ event into $(\ell jj)_1
(\ell jj)_2$ clusters, with invariant masses $m_1,\;m_2$,  in six different
ways; each such partition represents a possible $(\chi_1^0)(\chi_1^0)$
reconstruction; the best reconstruction is the one with least difference
between $m_1$ and $m_2$, and the best reconstructed $\chi_1^0$ mass is
the corresponding mean value $ m(\ell jj) = (m_1+m_2)/2$, with a
distribution peaked near the true value $m_{\chi_1^0}$. \\
(c) $U^cD^cD^c$-mediated decays: Eq.(\ref{eq:dec3}). Each $\chi_1^0$ decays
to three quarks (i.e. potentially three jets) with no missing energy.
There are 10 ways to partition 6 jets into two 3-jet clusters with masses
$m_1$ and $m_2$, say, to provide candidate reconstructions; as before,
the best reconstruction is the one with least difference between $m_1$
and $m_2$, and the best $\chi_1^0$ mass estimate is the mean value
$ m(jjj) = (m_1+m_2)/2$, with a distribution peaked near the true
value $m_{\chi_1^0}$.\\
(d)  Note finally that if $\lambda$ happens to lie in the range
$5\times (10^{-4}\mbox{--}10^{-6})\sqrt {\beta\gamma}\; (\tilde m_f)^2 \;
(m_{\chi_1^0})^{-5/2}$,  giving mean decay lengths of order
0.1\,mm--1\,m, the two decay vertices will usually be detectably displaced
from each other and from the beam-intersection spot, giving an important
additional signature.

The cross section for Eq.(\ref{eq:prod}) depends on the composition
of $\chi_1^0$ in terms of electroweak gauginos and higgsinos.
The production proceeds via $t$- and $u$-channel exchanges of selectrons and an
$s$-channel $Z$-pole. The couplings in the general case are
\begin{equation}
{\cal L} = e\sqrt 2 \, \bigl(  f_L \tilde e_L \bar e_L \chi_1^0
                             + f_R \tilde e_R \bar e_R \chi_1^0 + h.c.\bigr)
                  + e c_a Z^\mu \bar\chi_1^0 \gamma_\mu \gamma_5 \chi_1^0
\end{equation}
with the coefficients $f_L,\ f_R$ and $c_a$ determined by the gaugino and
higgsino composition of the neutralino. For pure states these coefficients take
the values
\begin{equation}
\begin{array}{lccccc}
\multicolumn{1}{c}{\chi} & & f_L & f_R & c_a\\
\hline
\rm photino & \tilde\gamma & 1 & 1 & 0\\
\rm zino & \tilde Z  & g_L^e & g_R^e & 0\\
\rm bino & \tilde\lambda_0& {1\over 2\cos\theta_w}& {1\over\cos\theta_w} & 0\\
\rm wino & \tilde\lambda_3 & {1\over2\sin\theta_w} & 0 & 0\\
\rm higgsino & \tilde h &   0 & 0 &  -{1\over4\sin\theta_w\cos\theta_w} \\
\rm higgsino & \tilde h'&   0 & 0 &   {1\over4\sin\theta_w\cos\theta_w} \\
\hline
\end{array}
\end{equation}
where $g_L^e=(1-2\sin^2\theta_w)/\sin2\theta_w$, $g_R^e=-\tan\theta_w$,
and $\theta_w$ is the usual weak angle.
The weak isospins  $T_3$ of $\tilde h$ and $\tilde h'$ are $1\over2$ and
$-{1\over2}$ respectively.
In general,
\begin{equation}
f_{L,R}=\sum_i f_{L,R}^i a_i \epsilon_{L,R}
\end{equation}
\begin{equation}
c_a={a_{\tilde h'}^2
- a_{\tilde h}^2
     \over
      4\sin\theta_w\cos\theta_w}
\end{equation}
where we denote the $\tilde h$, $\tilde h'$, $\tilde \lambda_3$,
$\tilde\lambda_0$ components of  $\chi_1^0$ by $a_i$.
The phase factor $\epsilon_{L}$ is 1 (or $i$) for a positive (or negative)
eigenvalue of of the state $\chi_1^0$ when the neutralino
mass matrix is diagonalized. Also $\epsilon_R=\epsilon_L^*$.
Our expression for $c_a$ agrees with Ref.\cite{BBDKT}.

The cross section is evaluated in the helicity basis $L,R$ for subsequent
use in joint production and decay analysis. The production amplitudes are
\begin{equation}
{\cal M}(e_L \bar e_R\to \chi_L\chi_R) = - e^2\cos^2{\theta\over2}
\Biggl[ f_L^2  \Biggl( {s(1-\beta)\over t-\tilde m_{eL}^2}
                      -{s(1+\beta)\over u-\tilde m_{eL}^2} \Biggr)
        + {4\beta c_a g_L^e s \over s-M_Z^2 + iM_Z \Gamma_Z} \Biggr]  \\
\end{equation}
\begin{equation}
{\cal M}(e_L \bar e_R\to \chi_R\chi_L) = - e^2\sin^2{\theta\over2}
\Biggl[ f_L^2  \Biggl( {s(1-\beta)\over u-\tilde m_{eL}^2}
                      -{s(1+\beta)\over t-\tilde m_{eL}^2} \Biggr)
        + {4\beta c_a g_L^e s \over s-M_Z^2 + iM_Z \Gamma_Z} \Biggr]  \\
\end{equation}
\begin{equation}
 {\cal M}(e_L\bar e_R\to\chi_L\chi_L)=
-{\cal M}(e_L\bar e_R\to \chi_R\chi_R)
= e^2f_L^2\sin\theta\;{m_{\chi_1^0}\over\sqrt s} \left
  ( {s\over t-\tilde m_{eL}^2} - {s\over u- \tilde m_{eL}^2}\right)
\end{equation}
\begin{equation}
{\cal M}(e_R \bar e_L\to \chi_R\chi_L) = \; e^2\cos^2{\theta\over2}
\Biggl[ f_R^2  \Biggl( {s(1-\beta)\over t-\tilde m_{eR}^2}
                      -{s(1+\beta)\over u-\tilde m_{eR}^2} \Biggr)
        - {4\beta c_a g_R^e s \over s-M_Z^2 + iM_Z \Gamma_Z} \Biggr]  \\
\end{equation}
\begin{equation}
{\cal M}(e_R \bar e_L\to \chi_L\chi_R) = \; e^2\sin^2{\theta\over2}
\Biggl[ f_R^2  \Biggl( {s(1-\beta)\over u-\tilde m_{eR}^2}
                      -{s(1+\beta)\over t-\tilde m_{eR}^2} \Biggr)
        - {4\beta c_a g_R^e s \over s-M_Z^2 + iM_Z \Gamma_Z} \Biggr]  \\
\end{equation}
\begin{equation}
 {\cal M}(e_R\bar e_L\to\chi_R\chi_R)=
-{\cal M}(e_R\bar e_L\to \chi_L\chi_L)
= e^2f_R^2\sin\theta\;{m_{\chi_1^0}\over\sqrt s} \left
  ( {s\over t-\tilde m_{eR}^2} - {s\over u- \tilde m_{eR}^2}\right)
\end{equation}
where $s,t,u$ are the usual invariant squares of CM energy and
momentum transfer, $\theta$ is the polar scattering angle, and
$\beta=\sqrt {1-4m_{\chi_1^0}^2/s}$ is the CM velocity of $\chi_1^0$.
Note that production from $e_L\bar e_R$ ($e_R\bar e_L$) initial
states involves $\tilde e_L$ ($\tilde e_R$) exchanges; we assume
no $\tilde e_L,\,\tilde e_R$ mixing, which is expected to be an
excellent approximation. The differential cross section is given by
\begin{equation}
{d\sigma\over d\cos\theta} = {\beta\over 128\pi s} \sum |{\cal M}|^2
\; .
\end{equation}
Since we have identical particles in the final state, the phase space is
limited to the forward hemisphere, $\cos\theta \ge 0$.
In the case that $\chi_1^0$ is a photino, we reproduce the cross section
expression given in Ref.\cite{ellis}.

The decay amplitude is greatly simplified if we assume it
is dominated by the exchange of the right-handed scalar partner.
For the process $\chi_h(p) \rightarrow \mu(q) e^+(l) \bar\nu_\tau(k)$,
the $R$-parity violating coupling $\lambda_{132}$ gives
\begin{equation}
{\cal M}={e\sqrt{2} f_R \lambda_{132}
      \over (l+k)^2-\tilde m_{\mu_R}^2 }
T_{+,h}(q,p) T_{-,+}(l,k)
\sqrt{8 l^0 k^0 q^0 (p^0- h |{\bf p}|)}  \ ,
\end{equation}
where $h=+$ or $-$ denotes the helicity, $R$ or $L$,  of the $\chi$
state; the spinor product $T$ is defined in Ref.\cite{HZ}.
We have assumed this form in our decay calculations.
Generalization of this formula  is straightforward.
The full amplitude to a given final state is the product of contributing
production and decay amplitudes summed over intermediate $\chi_1^0$
helicities; we have calculated distributions by Monte Carlo methods,
in the narrow-$\chi_1^0$-width approximation.

For our discussion, we focus particular attention on one interesting
example with  $m_{\chi_1^0}=48$ GeV, light sleptons
$\tilde m_{\ell R}=74$ GeV and $\tilde m_{\ell L}=112$ GeV
($\ell=e,\mu ,\tau$) plus relatively heavy squarks;
this is a typical SUSY-GUT solution from Ref.\cite{bbo}
in the top-Yukawa fixed-point region with $\tan\beta=1.5$,
where $\chi_1^0$ is almost a pure bino.  We also focus attention on
$\sqrt s =140$ GeV, a likely energy for intermediate LEP running, where
luminosity of order 20 pb$^{-1}$ may be accumulated in late 1995.

Figure 1(a) gives the integrated cross section for the particular
choice of energy and selectron masses above, but allowing the
$\chi_1^0$ mass and composition to vary. We see that of order twenty
$e^+e^-\to\chi_1^0\chi_1^0$ events could soon be produced at
$\sqrt s=140$ GeV if $\chi_1^0$ is light and bino-like (as in our
example above) or photino-like or higgsino-like.
Figure 1(b) illustrates the uncut angular distribution of
$\chi_1^0$ production for $m_{\chi_1^0}=48$~GeV, with the same choice
of energy and selectron masses as before.  For the interesting bino-like
and photino-like cases, we see that $\chi_1^0$ production is
preferentially at large polar angles $\theta$, away from the beam-pipe
region where detection is poor; however the higgsino-like case has
less favorable dependence $d\sigma /d\cos\theta\sim 1+\cos^2\theta$,
offsetting the higher integrated cross section here.
Signals from $LLE^c$ or $LQD^c$ decay modes can in principle
be detected by their exotic leptonic content (especially same-sign
dileptons) and generally have small SM backgrounds \cite{grt};
a handful of such events should be enough to attract
attention and provoke a detailed analysis.  Six-jet signals from
$U^cD^cD^c$ decays are less remarkable, since there are
$e^+e^-\to q\bar qgggg$ and other QCD backgrounds;
however, these backgrounds are expected to be of order
$\alpha^2 \alpha_s^4 /s$  which is small compared to our illustrated
signals of order $\alpha^2 \, s/ \tilde m_{eR}^4 \; \sim \alpha^2/s$,
and the signals contain a mass peak that we illustrate below.
All RPV signals could also contain displaced vertices,
as noted in (d) above.

To illustrate final-state invariant mass distributions, we
impose gaussian smearing on energies, with
$\Delta E/E = 0.8/\sqrt E$ for jets and
$\Delta E/E = 0.2/\sqrt E$ for leptons, to simulate
experimental resolution.  We also impose loose
semi-realistic cuts, requiring each of the visible leptons or
quarks (=jets) to have energy $E > 6$ GeV, rapidity $|\eta | < 2$
and angular separations $\Delta\theta_{ij} > 15^\circ$.
For $m_{\chi_1^0}=48$ GeV,
these cuts typically reduce the four-item (six-item) final
state event rates by factors 0.5 (0.3), where item means charged
lepton or jet.  We then take the same energy and selectron masses
as above, assume $\chi_1^0$ to be a pure bino, and compare the
case $m_{\chi_1^0}=48$ GeV (from the SUSY-GUT example\cite{bbo})
with the case $m_{\chi_1^0}=60$ GeV, to illustrate the mass sensitivity.

Figure 2(a) illustrates the distributions of unlike-sign
dilepton mass $m(\ell^+\ell^-)$ in four-lepton final states from
$LLE^c$ decays (four entries per event).  The signal is at best
a broad peak, because of the missing neutrino in each $\chi_1^0$
decay, but there is also an intrinsic background here from the $50\%$
of dilepton pairs that do not have the same $\chi_1^0$ parent.
Figure 2(b) shows the distribution of reconstructed $\chi_1^0$ mass
$m(\ell jj)$ in $\ell\ell jjjj$ final states from
$LQD^c$ decays; the background below the peak comes from
the gaussian energy resolution, which smears the peak and also leads to
some wrong choices for the best partitioning.
Figure 2(c) shows the distribution of reconstructed $\chi_1^0$ mass
$m(jjj)$ in $jjjjjj$ final states from $U^cD^cD^c$ decays (in the
limit of large exchanged squark mass); here too the
background outside the peak is due to energy smearing.
In all three of these distributions, we see there is sensitivity to
the $\chi_1^0$ mass; in the latter two cases there is a clear mass
peak showing high sensitivity.

To summarize, we have shown that the SUSY threshold process Eq.(1)
could be recognizable and detectable at pre-LEP200 energies such as
$\sqrt s=140$ GeV, with modest luminosities such as 20 pb$^{-1}$,
if $R$-parity is violated and $\chi_1^0$ is the LSP.
We have demonstrated that the production angular distribution
may well be favorable to detection (Fig.1(b)).
For the different possible RPV decay mechanisms, we have also
illustrated --- with semi-realistic energy resolution and acceptance
--- how appropriate final-state invariant mass distributions can be
used to extract the $\chi_1^0$ mass (Figs.2(a)--2(c)).
Our results are complementary to those of Ref.\cite{grt}, which also
addressed RPV signals from Eq.(1) but calculated for $\sqrt s=200$~GeV,
focusing on $\tau$-lepton channels from $LLE^c$ and $LQD^c$ decays,
with no discussion of production angles nor final state invariant mass
distributions nor possible displaced-vertex signatures.

\begin{flushleft}{\bf Acknowledgments}\end{flushleft}
This research was supported in part by the U.S.~Department of Energy
under Grants No.~DE-FG02-95ER40896 and No.~DE-FG02-84ER40173 and
in part by the University of Wisconsin Research Committee with funds
granted by the Wisconsin Alumni Research Foundation.


\section*{Figures}
\begin{enumerate}

\item{$e^+e^- \to \chi_1^0 \chi_1^0$ production cross sections:
(a) integrated cross section versus $\chi_1^0$ mass $m_{\chi_1^0}$,
and (b) CM differential cross section versus $|\cos\theta |$, for
various cases. Solid, dashed, dot-dashed, short-long-dashed and
short-short-long-dashed curves denote the cases where $\chi_1^0$ is
purely photino, bino, zino, neutral wino and higgsino, respectively.
We here take CM energy $\sqrt s=140$ GeV with exchanged selectron
masses $\tilde m_{eL} = 112$ GeV and $\tilde m_{eR} = 74$ GeV.
\label{fig:fig1}}

\item{Invariant mass distributions that can reveal the $\chi_1^0$
mass $m_{\chi_1^0}$: (a) opposite-sign dilepton mass $m(\ell^+\ell^-)$
in four-lepton signals from $LLE^c$ decays;
(b) best-reconstructed lepton-plus-dijet mass $ m(\ell jj)$
in two-lepton-plus-four-jet signals from $LQD^c$ decays;
(c) best-reconstructed trijet mass $ m(jjj)$
in six-jet signals from $U^cD^cD^c$ decays. We assume
the same energy and selectron masses as in Fig.1, and compare
the cases $m_{\chi_1^0}=48,\;60$ GeV for a pure bino $\chi_1^0$.
\label{fig:fig2}}

\end{enumerate}

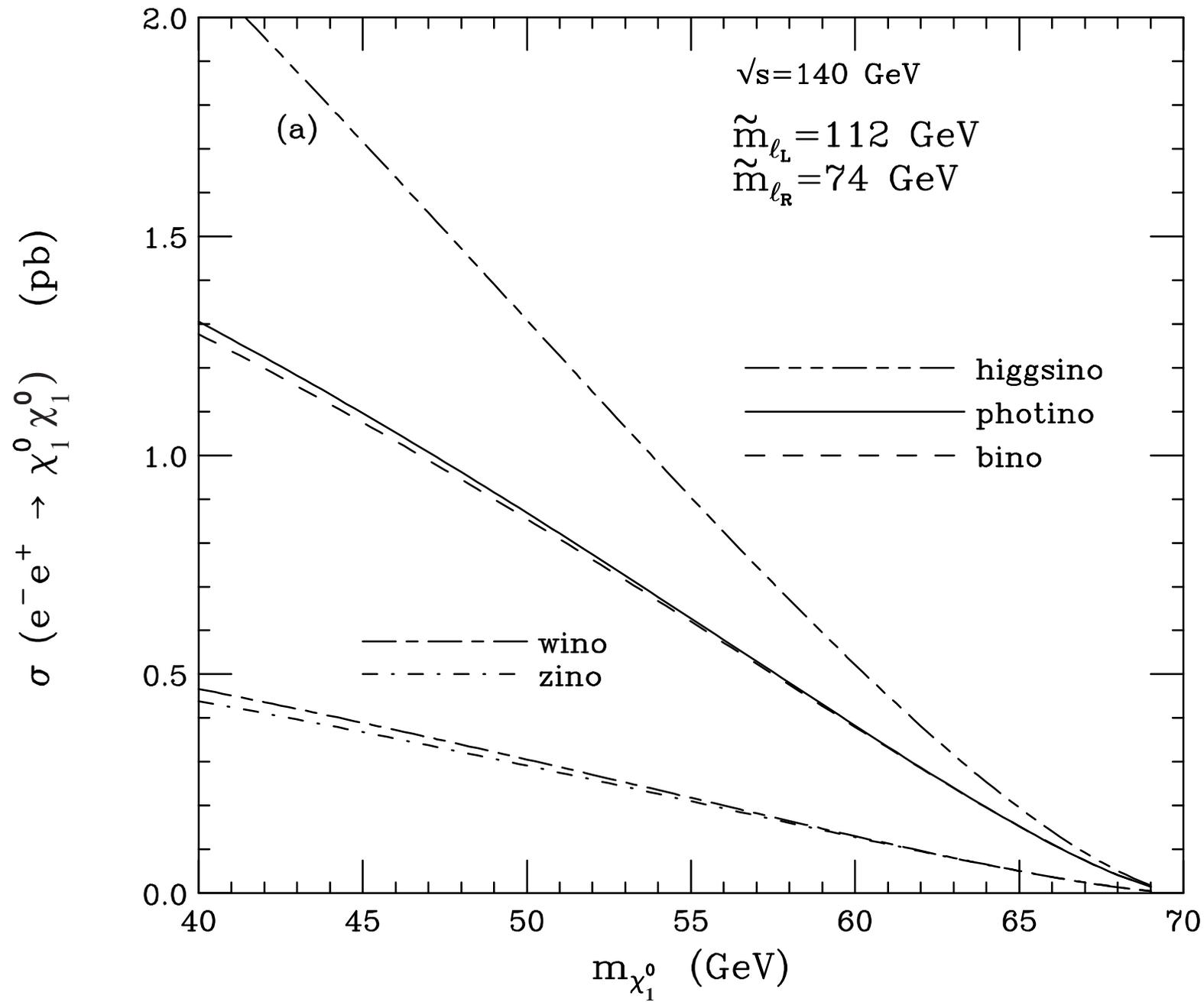
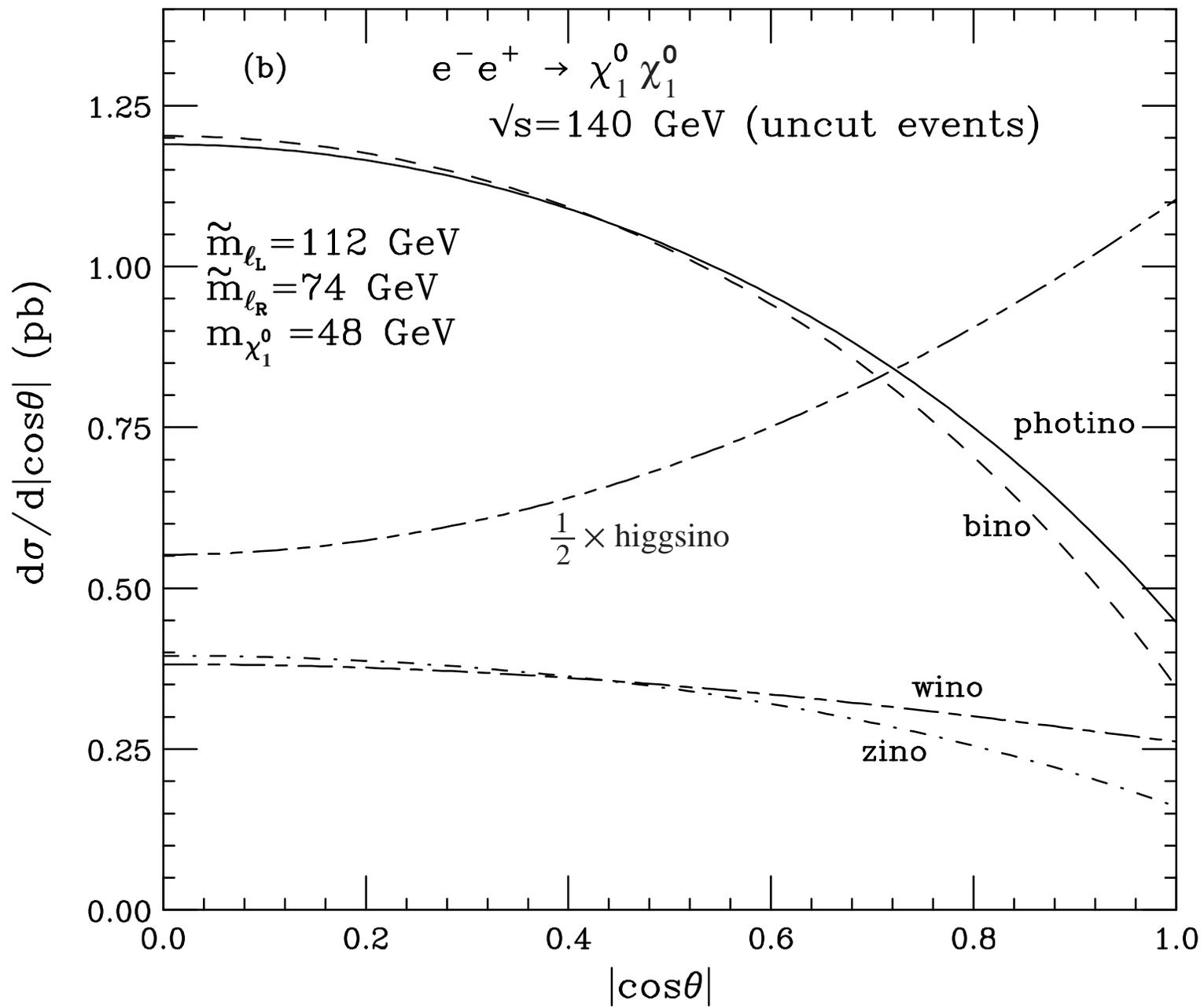
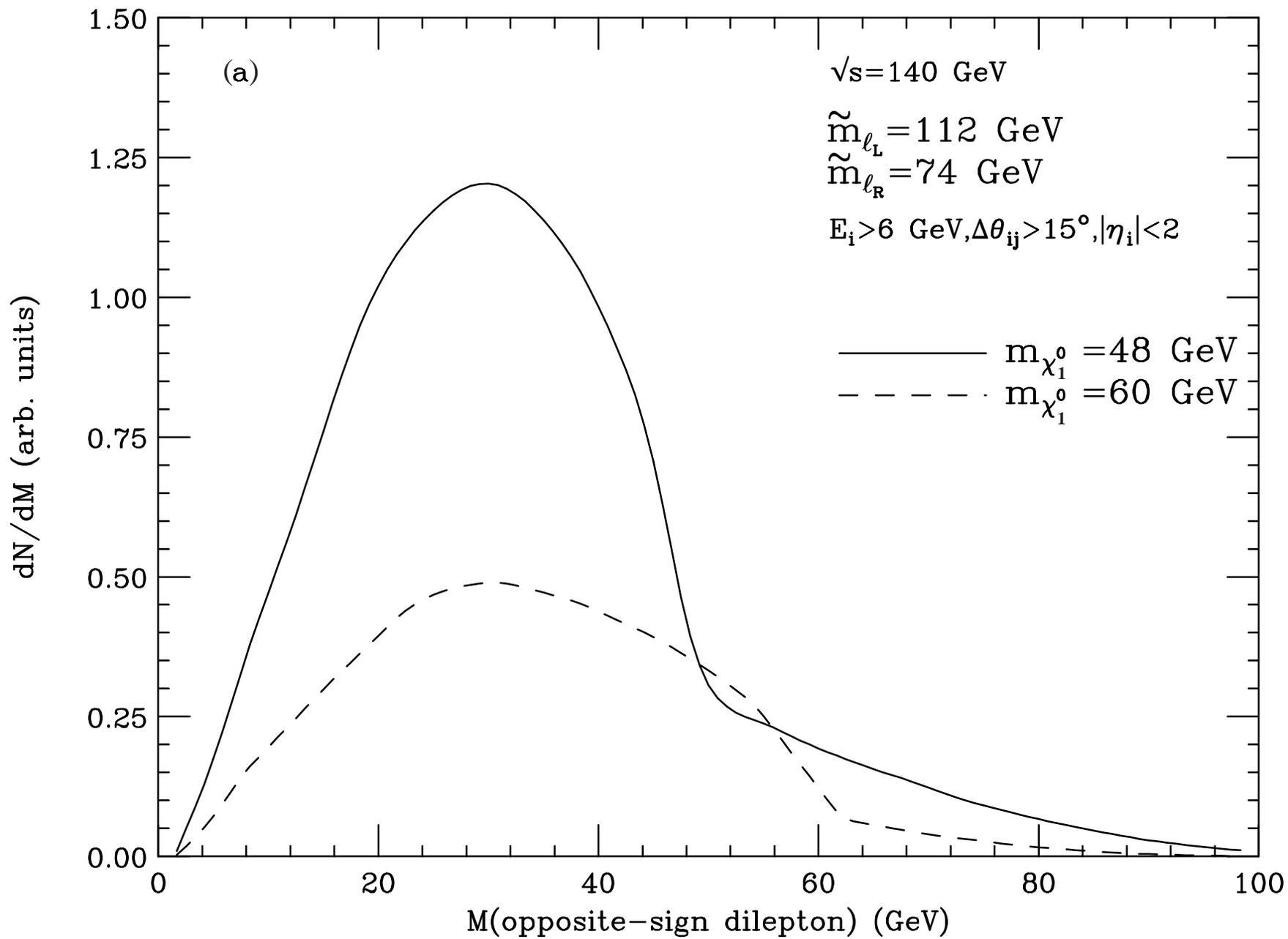
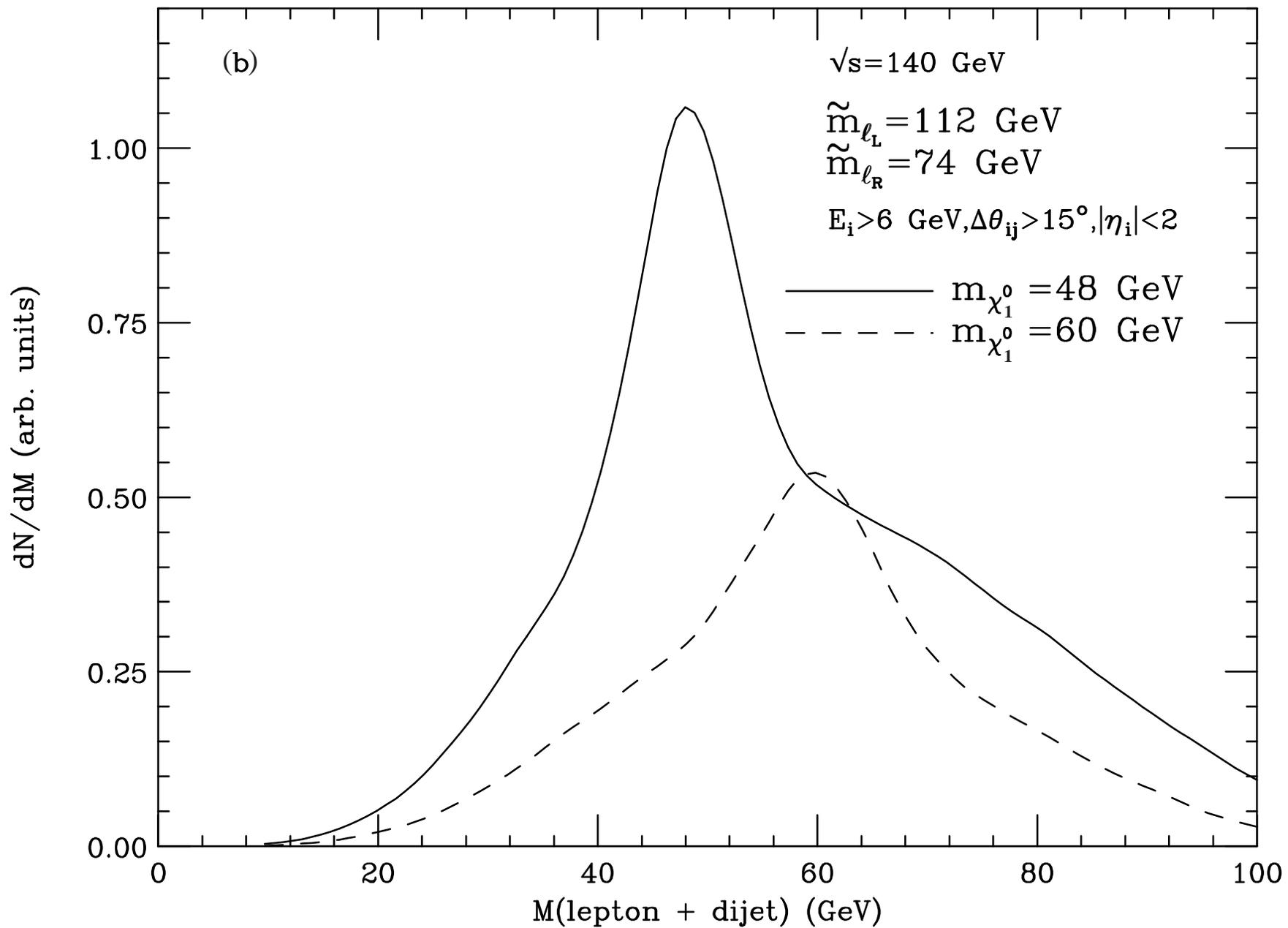
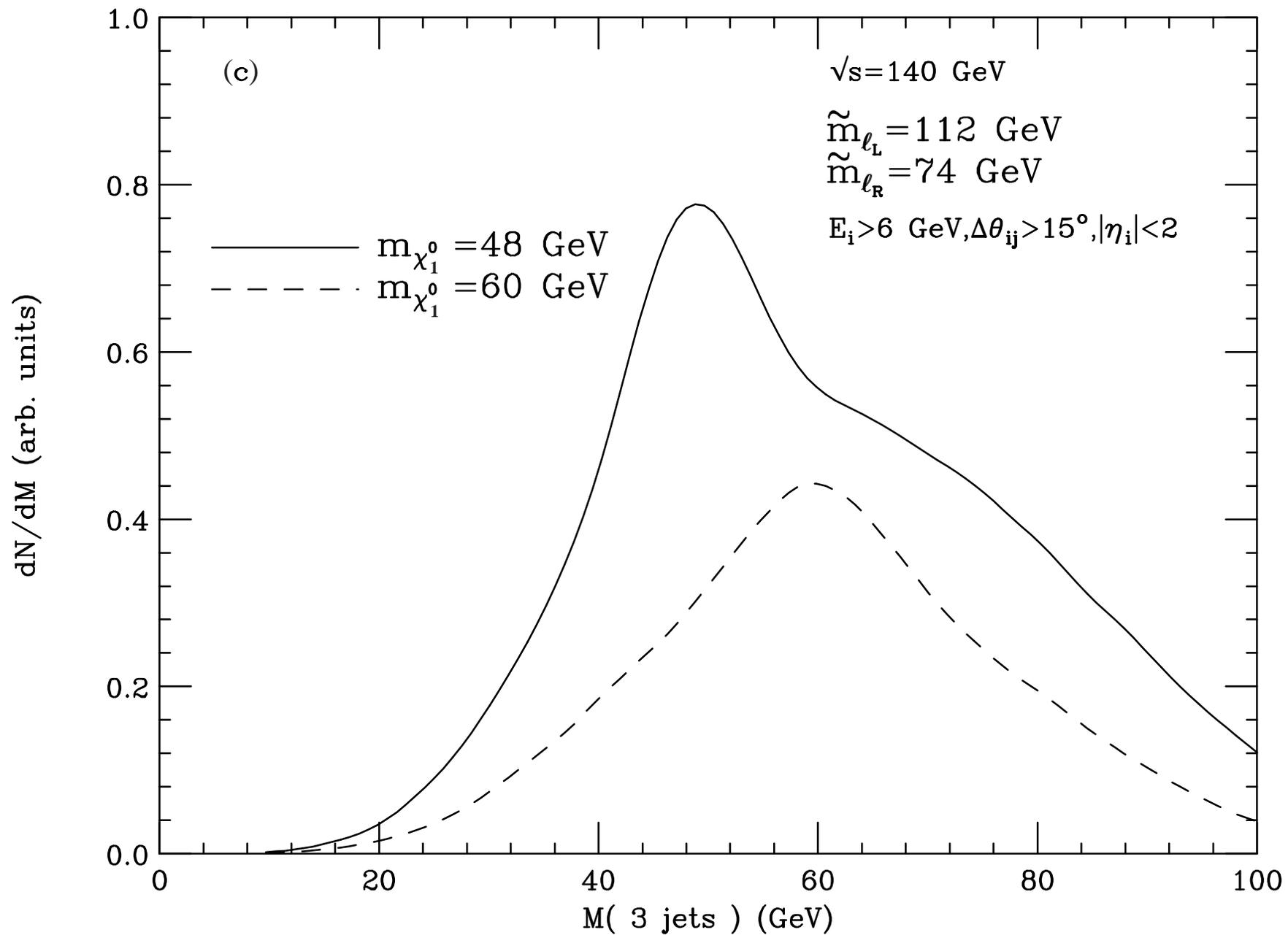
\end{document}